\documentclass[12pt]{article}
\usepackage{hyperref}
\hypersetup{
colorlinks=true,
linkcolor=black,
citecolor=black,
urlcolor=cyan,
filecolor=black,
pdfborder={0 0 0},
}

\usepackage{amssymb}
\usepackage{amsfonts}
\usepackage{mathrsfs}
\usepackage{latexsym}
\usepackage{amsmath}
\usepackage{graphics}
\usepackage{graphicx}
\usepackage{sectsty}
\usepackage{setspace}
\usepackage{multirow}
\sectionfont{\normalsize\bf}
\subsectionfont{\rm\normalsize}
\def\be{\begin{equation}}
\def\ee{\end{equation}}
\input amssym.def
\input amssym
\def\O{{\cal O}}
\def\A{{\cal A}}
\def\R{{\cal R}}
\def\L{{\cal L}}
\def\YM{{\hbox{\vphantom{g}\tiny YM}}}
\def\half{{\scriptstyle {1\over 2}}}

\def\Pf{\hbox{Pf }}

\numberwithin{equation}{section}

\begin{document}

\thispagestyle{empty}
\begin{flushright}
\end{flushright}
\baselineskip=16pt
\vspace{.5in}
{
\begin{center}
{\bf Proof of the Formula of Cachazo, He and Yuan}

{\bf for Yang-Mills Tree Amplitudes in Arbitrary Dimension}
\end{center}}
\vskip 1.1cm
\begin{center}
{Louise Dolan}
\vskip5pt

\centerline{\em Department of Physics}
\centerline{\em University of North Carolina, Chapel Hill, NC 27599} 
\bigskip
\bigskip        
{Peter Goddard}
\vskip5pt

\centerline{\em School of Natural Sciences, Institute for Advanced Study}
\centerline{\em Princeton, NJ 08540, USA}
\bigskip
\bigskip
\bigskip
\bigskip
\end{center}

\abstract{\noindent 
A proof is given of the formula, recently proposed by Cachazo, He and Yuan  (CHY)  for gluon tree amplitudes in pure
Yang-Mills theory in arbitrary dimension. The approach is to first establish the corresponding  result for massless
$\phi^3$ theory using the BCFW recurrence relation and then to extend this to the gauge theory case.
 Additionally, it is shown that the scattering equations introduced by CHY can be generalized to massive particles, enabling the description of tree amplitudes for massive
$\phi^3$ theory.
}
\bigskip

\setlength{\parindent}{0pt}
\setlength{\parskip}{6pt}

\setstretch{1.05}
\vfill\eject
\vskip50pt
\section{\bf Introduction}

Recently Cachazo, He and Yuan (CHY) have proposed a compact and relatively simple formula for the sum of all $N$-point tree diagrams for pure Yang-Mills theory in an arbitrary dimension of space-time and a similar formula for the sum of all $N$-point  tree diagrams for gravity \cite{CHY1}. These formulae express the $N$-point amplitudes as sums over the solutions for $z_a$ to the equations 
\be
\hat f_a(z,k)=0,\quad a\in A,\qquad \hbox{where}\quad \hat f_a(z,k)=\sum_{b\in A\atop b\ne a}{k_a\cdot k_b\over z_a-z_b},\label{SE}
\ee
and where $k_a$ are the momenta of the particles labeled by $a\in A$, with $N=|A|$, which CHY call the scattering equations.  It follows from momentum conservation, $\sum_{a\in A}k_a=0$, and the condition that the particles are massless, $k_a^2=0, a\in A$, that  this system of equations is invariant under M\"obius transformations,
\be 
z_a\mapsto \zeta_a={\alpha z_a+\beta\over\gamma z_a+\delta},\quad a\in A,\label{Mt}
\ee
in that, if $z=(z_a)$ is a solution to (\ref{SE}), $\zeta=(\zeta_a)$ also provides a solution: $\hat f_a(\zeta,k)=0, a\in A$. 

Because of this symmetry under the three-dimensional M\"obius group, only $N-3$ of the equations (\ref{SE}) are independent, and we can restrict to their solutions by means of the delta functions
\be
{\prod_{a\in A\atop }}'\delta\left(\hat f_a(z,k)\right)\equiv (z_i-z_j)(z_j-z_k)(z_k-z_i)\prod_{a\in A\atop a\ne i,j,k}\delta\left(\hat f_a(z,k)\right)\label{prodp}
\ee
which is independent of the choice of $i,j,k\in A$. Under the M\"obius transformation (\ref{Mt}), 
\be
{\prod_{a\in A\atop }}'\delta\left(\hat f_a(z,k)\right)\mapsto {\prod_{a\in A\atop }}'\delta\left(\hat f_a(\zeta,k)\right)=\prod_{a\in A\atop }{\alpha\delta-\beta\gamma\over (\gamma z_a+\delta)^2}{\prod_{a\in A\atop }}'\delta\left(\hat f_a(z,k)\right),
\ee
so that the integrand of 
\be
\A_N=\int \widehat\Psi_N(z;k;\epsilon){\prod_{a\in A\atop }}'\delta\left(\hat f_a(z,k)\right)\prod_{a\in A}{dz_a\over (z_a-z_{a+1})^2}\bigg/d\omega\label{amp1}
\ee
is M\"obius invariant provided that the function $\widehat\Psi_N(z;k;\epsilon)$ is itself M\"obius invariant. In (\ref{amp1}), we have chosen a cyclic ordering on $A$, with $a+1$ indicating the label next after $a$ in the ordering, say $(1,2,\ldots,N)$ with $N+1\equiv 1$, $\widehat\Psi_N(z;k;\epsilon)$ may depend on the momenta $k_a$ and polarizations,  $\epsilon_a$, if any, of the particles, and $d\omega$ denotes the invariant measure on the M\"obius group,
\be
d\omega={dz_rdz_sdz_t\over (z_r-z_s)(z_s-z_t)(z_t-z_r)}\;. \label{Mm}
\ee
In \cite{CHY1}, CHY propose, and present evidence for, an explicit form for $\widehat\Psi_N(z;k;\epsilon)$ for the sum of $N$-gluon tree diagrams in pure gauge theory, in an arbitrary dimension of space-time, for a given order of the external gluons (to be multiplied by an appropriate color factor). 
In order to give a precise interpretation of (\ref{amp1}) that includes a sum over all, possibly complex, solutions of the  equations (\ref{SE}),  CHY rewrite the expression  as a contour integral 
\be
\A_N=\oint_\O \widehat\Psi_N(z;k;\epsilon){\prod_{a\in A\atop }}'{1\over\hat f_a(z,k)}\prod_{a\in A}{dz_a\over (z_a-z_{a+1})^2}\bigg/d\omega\label{amp}
\ee
where the contour $\O$ encloses all the solutions of (\ref{SE}) and ${\prod}'$ is defined as in (\ref{prodp}).

Here, following the approach we used in \cite{DG}, 
to establish the complete equivalence of  gluon tree amplitudes in open twistor string theory \cite{BW} to the corresponding amplitudes in $N=4$ super Yang-Mills theory, we prove these CHY formulae using the BCFW relations \cite{BCFW}, by first showing that taking $\widehat\Psi_N(z;k;\epsilon)$ constant in (\ref{amp1}) gives the tree amplitudes for massless $\phi^3$ theory. [In \cite{CHY2}, CHY have also noted that taking $\widehat\Psi_N(z;k;\epsilon)$ constant in (\ref{amp1}) can provide the amplitudes for a scalar theory, but they consider a theory with $U(N)\times U(\tilde N)$ symmetry.]

The CHY equations $\hat f_a(z,k)=0, a\in A, $ can be modified to describe massive particles  (although we only need the massless case as a precursor to our proof of the CHY proposal for gauge theory tree amplitudes).  Having chosen a cyclic ordering on $A$, we consider
\be
\check f_a(z,k)=\sum_{b\in A\atop b\ne a}{k_a\cdot k_b\over z_a-z_b}+{m^2\over 2(z_a-z_{a+1})}+{m^2\over 2(z_a-z_{a-1})},\qquad a\in A,\label{MSE}
\ee
where, in the signature we are using, $k_a^2=m^2, a\in A$. This system of equations is again M\"obius invariant, and now (\ref{amp}) produces the tree amplitudes for a $\phi^3$ theory for a scalar particle of mass $m$. However, unlike the original, these modified CHY equations depend on the order of particles which previously only entered through what one might regard as the measure factor in the integrand of (\ref{amp1}) or (\ref{amp}).

After reviewing the general properties of the CHY scattering equations in section \ref{SEqs}, we give a calculation in this formalism of the elementary cases of the $4-$ and $5$-point amplitudes in massless $\phi^3$ in section \ref{phi31}, noting that if we have chosen $z_1\rightarrow\infty$, the formulae would apply even when $k_1$ is off-shell.  In section \ref{phi32}, we give a direct inductive proof of the formula for $\phi^3$ theory, using a form of the result that works when one of the momenta is off shell. This approach would be very cumbersome for gauge theory, so we give a second proof in section \ref{phi33} using the BCFW relations. This analysis is extended in section \ref{GT} to give a proof of the formula of CHY \cite{CHY1} for the tree amplitudes of pure gauge theory. 
Results on Pfaffians, necessary for the discussion of section \ref{GT} are described in appendix \ref{Pfaff}.  The extension of the results on $\phi^3$ theory to the massive case is discussed in appendix \ref{massive}.
\vfil\eject
\section{\bf Properties of the CHY Scattering Equations}
\label{SEqs}
We review some properties of the CHY scattering equations (\ref{SE}). 

{\sl M\"obius Invariance.}  This system of equations  is M\"obius invariant provided that momentum is conserved, {\it i.e.} $\sum_{a\in A}k_a=0$, and all the particles are massless, {\it i.e.}
$k_a^2=0, a\in A$.
To see this note that
\be
\sum_{b\in A\atop b\ne a}{k_a\cdot k_bz_b\over z_a-z_b}=-\sum_{b\in A\atop b\ne a}{k_a\cdot k_b}+z_a\sum_{b\in A\atop b\ne a}{k_a\cdot k_b\over z_a-z_b}=k^2_a,\quad a\in A,\label{zf}
\ee
so that, 
\be\hbox{if }\zeta_a={\alpha z_a+\beta\over\gamma z_a+\delta},\quad \sum_{b\in A\atop b\ne a}{k_a\cdot k_b\over \zeta_a-\zeta_b}
={\gamma z_a+\delta\over\alpha\delta-\beta\gamma}\sum_{b\in A\atop b\ne a}{k_a\cdot k_b(\gamma z_b+\delta)\over z_a-z_b}=0,\label{k2}\ee
provided that $k_a^2=0$ for all $a\in A$. Thus, then the system of $N$ equations (\ref{SE}) determines the $N$ variables $z_a$ only up to M\"obius  transformation, implying only $N-3$ of them are independent and there must be three relations between them.  The equations determine the $N-3$ independent cross-ratios of the $z_a$; we are free to specify any three of the $z_a$ as we wish and the others are then determined by (\ref{SE}).

For general values of $z$, 
 the functions $\hat f_a(z,k)$ satisfy the relations:
\be
\sum_{a\in A\atop }\hat f_a(z,k)=\sum_{a,b\in A\atop b\ne a}{k_a\cdot k_b\over z_a-z_b}=0,
\label{sumf}
\ee
independently of the momentum conservation and zero-mass conditions on the $k_a$;
\be
\sum_{a\in A\atop }z_a\hat f_a(z,k)=\sum_{a,b\in A\atop b\ne a}{z_ak_a\cdot k_b\over z_a-z_b}={1\over 2}\sum_{a,b\in A\atop b\ne a}k_a\cdot k_b={1\over 2}\left[\sum_{a\in A}k_a\right]^2=0,\label{c12}\ee
using first the zero-mass conditions, $k_a^2=0$, and then momentum conservation; and
\be\sum_{a\in A\atop }z_a^2\hat f_a(z,k)={1\over2}\sum_{a,b\in A\atop b\ne a}{(z_a^2-z_b^2)k_a\cdot k_b\over z_a-z_b}={1\over2}\sum_{a,b\in A\atop b\ne a}{(z_a+z_b)k_a\cdot k_b}=-\sum_{a\in A\atop }z_ak^2_a=0.
\label{c3}\ee
From (\ref{c12}) and (\ref{c3}), we have that
\be
\left|{\partial(\hat f_a,\hat f_b,\hat f_c)\over\partial(\hat f_i,\hat f_j,\hat f_k)}\right|=\left|\begin{matrix} 1&1&1\cr z_i&z_j&z_k\cr z_i^2&z_j^2&z_k^2\end{matrix}\right|\times\left|\begin{matrix} 1&1&1\cr z_a&z_b&z_c\cr z_a^2&z_b^2&z_c^2\end{matrix}\right|^{-1}={(z_i-z_j)(z_j-z_k)(z_k-z_i)\over (z_a-z_b)(z_b-z_c)(z_c-z_a)}
\ee
(\ref{prodp})
when $\hat f_d(z,k)=0$ for $d\ne a,b,c,i,j,k$, which establishes that (\ref{prodp}) is indeed independent of the choice of $i,j,k\in A$.

{\sl Relation to  tree amplitude integrands in string theory.} The $N$-point tree amplitude in the original bosonic open string theory is given, in Koba-Nielsen variables, by an integral
\be
A_N^{\hbox{\tiny string}}=\int V(z;k)\prod_{a\in A}{dz_a\over z_a-z_{a+1}}\bigg/d\omega\label{samp}
\ee
with
\be
V(z,k)=\prod_{1\leq a<b\leq N}(z_a-z_b)^{-\alpha'k_a\cdot k_b}\prod_{a=1}^N(z_a-z_{a+1})^{\alpha_0}\equiv U(z,k)^{\alpha'},
\ee
where the leading trajectory, $\alpha(s)=\alpha_0+\half \alpha's$, with intercept $\alpha_0$ and slope parameter $\alpha'$, and the particles satisfy the mass condition $k_a^2=m^2$, where $m^2=-2\alpha_0/\alpha'=-2a_0$.
[Scherk obtained $\phi^3$ theory by taking $\alpha'\rightarrow 0$, at fixed $a_0$, in (\ref{samp}) in his original paper on the zero-slope (infinite-tension)
limit \cite{S1}. In this connection see the recent papers by Mason and Skinner \cite{MS} and by Berkovits \cite{B}.]

Crucially, $V(z,k)$, or equivalently,
\be
U(z,k)=\prod_{1\leq a<b\leq N}(z_a-z_b)^{-k_a\cdot k_b}\prod_{a=1}^N(z_a-z_{a+1})^{a_0},
\ee
is M\"obius invariant, that is, if $\zeta_a$ is given by (\ref{k2}), $U(\zeta,k)=V(z,k)$. Then, if we define $\breve f_a(z,k)$ by 
\be
{\partial U(z,k)\over \partial z_a}=-\breve f_a(z,k) U(z,k),
\ee
then $\breve f_a(z,k)$ is given by (\ref{MSE}) and the M\"obius invariance of $U(z,k)$ implies 
\be
\breve f_a(\zeta,k) ={\partial z_a\over \partial \zeta_a}\breve f_a(z,k)={(\gamma z_a+\delta)^2\over(\alpha\delta-\beta\gamma)^2}\breve f_a(z,k),
\ee
so that the set of equations $\breve f_a(z,k)=0, a\in A,$ is M\"obius invariant. $\breve f_a(z,k)=\hat f_a(z,k)$ when $a_0=0$.

The M\"obius invariance of $U(z,k)$ immediately implies the relations (\ref{sumf})-(\ref{c3}), because, for an infinitesimal M\"obius transformation $\delta z_a=\epsilon_1+\epsilon_2 z_a+\epsilon_3z_a^2$, 
\be 
\delta U=\sum_{a\in A}\delta z_a {\partial U(z,k)\over \partial z_a}=0,\qquad \hbox{implying}\qquad \sum_{a\in A}\delta z_a \breve f_a=0,
\ee
that is,
\be 
\sum_{a\in A}\breve f_a(z,k)=0;\qquad\sum_{a\in A}z_a\breve f_a(z,k)=0;\qquad\sum_{a\in A}z_a^2\breve f_a(z,k)=0.
\ee

{\sl Factorization.} Consider the situation in which, as the $k_a, a\in A$, vary in some specific way maintaing the zero-mass conditions $k_a^2=0$, two or more of the $z_a$ to tend to the same value, $z_S$, say. Specifically, suppose $z_a= z_S +\epsilon x_a+\O(\epsilon^2)$, as $\epsilon\rightarrow 0$, for $a\in S$, where $S\subset A$, and $z_a\not\rightarrow z_S$ for $a\notin S$. Then 
\be
\hat f_a(z,k)={1\over\epsilon}\hat g_a(x,k)\left[1+\O(\epsilon)\right],\quad\hat g_a(x,k)=\sum_{b\in S\atop b\ne a}{k_a\cdot k_b\over  x_a- x_b},\quad a\in S, \label{SES}
\ee
\be
\hat f_a(z,k)=\sum_{b\in S\atop }{k_a\cdot k_b\over z_a-z_S}+\sum_{b\notin S\atop b\ne a}{k_a\cdot k_b\over z_a-z_b}+\O(\epsilon),\quad a\notin S. \label{SES1}
\ee
Then, as in (\ref{sumf}),
\be
\sum_{a\in S} \hat g_a(x,k)=0,\label{sumg}
\ee
implying that if $\hat g_a=0$ for $a\in S, a\ne s$ ({\it e.g.} as the limit of conditions $\hat f_a=0$ for $a\in S, a\ne s$, as $\epsilon\rightarrow 0$), where $s$ is some particular element of $S$, then it follows that $\hat g_s(x)=0$ as well. Further,
multiplying (\ref{SES}) by $x_a-x_r$, for a choice of $r\in S$, and summing over $a\in S$, as in  (\ref{c12}), gives
\be
\sum_{a\in S}(x_a-x_r)\hat g_a(x,k)=\sum_{a,b\in S\atop b\ne a}{x_a-x_r\over  x_a- x_b}k_a\cdot k_b={1\over 2}\sum_{a,b\in S\atop b\ne a}k_a\cdot k_b=\half k_S^2,\qquad  k_S=\sum_{b\in S\atop }k_b. \label{SES2}
\ee
Then, if $\hat g_a(x,k)=0, a\in S, a\ne r,s$, we have that
\be
(x_s-x_r)\hat g_s(x,k)=\half k_S^2=(x_r-x_s)\hat g_r(x,k),\label{grs}
\ee
and, if additionally $g_r(x,k)=0$, then $k_S^2=0$. 

Imposing the equations $\hat f_a(z,k)=0,$ for all $a\in A,$ and taking the limit $\epsilon\rightarrow0$,
\be
\sum_{b\in S\atop b\ne a}{k_a\cdot k_b\over  x_a- x_b}=0,\quad a\in S;\qquad 
{k_a\cdot k_S\over z_a-z_S}+\sum_{b\notin S\atop b\ne a}{k_a\cdot k_b\over z_a-z_b}=0,\quad a\notin S;\qquad k_S^2=0.\label{kS2}
\ee
So, in this limit, the CHY equations (\ref{SE}) have in effect factored into two sets: the first set of equations are the CHY equations for the momenta $(k_a, a\in S; -k_S)$, with associated variables $(x_a, a\in S; \infty)$; and the second set are the equations for the momenta $(k_S;k_a, a\notin S)$, with associated variables $(z_S;z_a, a\notin S)$.

{\sl Specialization.} In what follows, in general it will be convenient to take $(r,s,t)$ in (\ref{Mm}) to be the same as $(i,j,k)$ in (\ref{prodp}) and further to take $i=1,j=2$ and $k=N$. Then (\ref{amp}) becomes
\be
(z_2-z_N)^2\int_\O \widehat\Psi_N(z;k;\epsilon)\prod_{a=3 }^{N-1}{dz_a\over\hat f_a(z,k)}\prod_{a=2 }^{N-1}{1\over (z_a-z_{a+1})^2}\label{amp2},
\ee
and making the particular choice $z_1=\infty, z_2=1, z_N=0$, this becomes
\be
\A_N=\oint_\O \widehat\Psi_N^o(z;k;\epsilon)\prod_{a=3 }^{N-1}{dz_a\over\hat f_a^o(z,k)}\prod_{a=2 }^{N-1}{1\over (z_a-z_{a+1})^2}\label{amp3},
\ee
where $\widehat\Psi_N^o(z;k;\epsilon)$ denotes the specialization of $\widehat\Psi_N(z;k;\epsilon)$ to these values and
\be
\hat f_a^o(z,k)={k_a\cdot k_2\over z_a-1}+\sum_{b=3\atop b\ne a}^{N-1}{k_a\cdot k_b\over z_a-z_b}+{k_a\cdot k_N\over z_a},\qquad 3\leq a\leq N-1.
\ee
To be clear about the position of singularities, it is convenient to rewrite this by setting
\be
f_a(z)=\hat f_a^o(z,k)\prod_{c=2\atop c\ne a}^N(z_a-z_c)=\sum_{b=2\atop b\ne a}^Nk_a\cdot k_b\prod_{c=2\atop c\ne a,b}^N(z_a-z_c),\label{fa}
\ee
\be
\A_N=-(-1)^{\half N(N-1)}\oint_\O{\widehat\Psi_N^o(z;k;\epsilon)\over (1-z_3)z_{N-1}}\prod_{a=3}^{N-2}z_a\prod_{a=4}^{N-1}(1-z_a)\prod_{b=5}^{N-1}\prod_{a=3}^{b-2}(z_a-z_b)^2\prod_{a=3}^{N-1}{dz_a\over f_a}.\label{ampsp}
\ee

\section{\bf  Massless ${\boldsymbol\phi}^{\bold 3}$ Theory}
\label{phi31}

In this section we discuss (\ref{amp}) in the simplest case, in which $\widehat\Psi_N(z;k;\epsilon)$ is constant, showing how, for $N=4,5$, (\ref{ampsp}) produces the amplitudes of massless $\phi^3$ theory in a form in which the mass-shell condition $k_1^2=0$ can be relaxed. In the next section, we will give an inductive proof of this partly off-shell formula for general $N$. Specifically, taking $\widehat\Psi_N(z;k;\epsilon)=(-2)^{3-N}$ in (\ref{amp}) and (\ref{ampsp}):
\begin{align}
\A_N^\phi&={1\over (-2)^{N-3}}\oint_\O {\prod_{a\in A\atop }}'{1\over\hat f_a(z,k)}\prod_{a\in A}{dz_a\over (z_a-z_{a+1})^2}\bigg/d\omega\label{ampphi1}\\
&={\varepsilon_N\over 2^{N-3}}\oint_\O{1\over (1-z_3)z_{N-1}}\prod_{a=3}^{N-2}z_a\prod_{a=4}^{N-1}(1-z_a)\prod_{b=5}^{N-1}\prod_{a=3}^{b-2}(z_a-z_b)^2\prod_{a=3}^{N-1}{dz_a\over f_a} ,\label{ampphi2}
\end{align}
where $\varepsilon_N=(-1)^{\half N(N+1)}$, $f_a$ is given by (\ref{fa}), the contour $\O$ encircles the simultaneous solutions of $f_a(z,k)=0, \;3\leq a\leq N-1$ (but excludes $z_3=1$ and $z_{N-1}=0$), and we need to sum $\A_N^\phi$ over the different cyclic orderings of the external legs. 

For $N=4$, writing $z=z_3$, this gives
\begin{align}
\A_4^\phi&={1\over 2}\oint_\O{dz\over (1-z)z[k_3\cdot k_2z+k_3\cdot k_4(z-1)]} ,\cr
&={k_3\cdot (k_2+ k_4)\over 2k_3\cdot k_2\; k_3\cdot k_4}={1\over 2k_3\cdot k_4}+{1\over 2k_3\cdot k_2}={1\over s}+{1\over t},
\label{amp4}
\end{align}
where $s=(k_3+k_4)^2, t=(k_2+k_3)^2,u=(k_2+k_4)^2$. Summing over the 6 different cyclic orderings (by permuting 2,3,4), and multiplying by $g^2/4$, where $g$ is the coupling constant,  gives the total amplitude $\A^{\phi\,\hbox{\tiny tot}}_4=g^2/s+g^2/t+g^2/u$ (because each term occurs four times), the appropriate result for $\phi^3$ theory, $\L=\half\partial^\mu\phi\partial_\mu\phi+(g/3!)\phi^3$. 

For the $N$-point function, $\A^{\phi\,\hbox{\tiny tot}}_N$, we need to multiply (\ref{ampphi2}) by  $(g/2)^{N-2}$ as we sum over the possible different orderings keeping $1$ fixed, because, as we shall prove, $\A_N^\phi$ gives the sum of the tree diagrams in $\phi^3$ theory that can be drawn in planar form for a given ordering of the external momenta [as $1/s$ in (\ref{amp4}) can be for the ordering $(1,2,3,4)$ but not the ordering $(1,3,2,4)$]. Each tree diagram will contribute to $2^{N-2}$ orderings, treating orderings related by cyclic rotations as equivalent. (If we treat orderings related by anti-cyclic or cyclic rotations as equivalent, we sum over half as many orderings and each diagram occurs $2^{N-3}$ times, so, in that case, we multiply by $g^{N-2}/2^{N-3}$.) This is the same counting as in Scherk's original discussion of the zero-slope limit of what became string theory \cite{S1}. It would be good to be able to derive (\ref{ampphi2}) directly from (\ref{amp}) by taking $\alpha'\rightarrow0$.  (See also \cite{MS} and \cite{B}.)  [We have chosen the sign of $\Psi_N$ so that $\A_N^\phi$ will equal a sum of tree diagrams with propagators $1/k^2$, where $k$ is the momentum. With this normalization, $\A^{\phi\,\hbox{\tiny tot}}_N$ equals  $i(-1)^N$ times the amplitude defined using the usual field theory convention.]

For $N=5$, writing $z=z_3, w=z_4$, we have
\be
\A_5^\phi=-{1\over 4}\oint_\O{z(1-w) dzdw\over (1-z)w f_3f_4},\label{ampN5}
\ee
where
\begin{align}
f_3&=k_3\cdot k_2z(z-w) +k_3\cdot k_4z(z-1)+k_3\cdot k_5(z-1)(z-w),\\
f_4&=k_4\cdot k_2w(w-z) +k_4\cdot k_3w(w-1)+k_4\cdot k_5(w-1)(w-z).\
\end{align}
This is most conveniently evaluated using the global residue theorem. We write 
\be
\R(\alpha,\beta)=-\lim_{\epsilon,\delta\rightarrow 0}{1\over 4}\oint_{\O_{\alpha\beta}}{z(1-w) dzdw\over (1-\epsilon-z)(w-\delta) f_3f_4}
\ee
for the residue of the integrand at $\alpha=\beta=0$, where $\O_{\alpha\beta}$  is an appropriately oriented contour about $\alpha=\beta=0$; $\alpha, \beta$ are chosen from the factors in the denominator of the integrand; and 
we have displaced the factors $1-z$ and $w$ by $\epsilon$ and $\delta$, respectively, in order to avoid singular configurations of these factors. Then
\be
\A_5^\phi=\R(f_3,f_4)=-\R(w,f_4)-\R(w,1-z)-\R(f_3,1-z)
\ee
The term $\R(w,1-z)$, for which the displacements by $\epsilon$ and $\delta$ are unnecessary, is easily evaluated:
\be
\R(w,1-z)=-{1\over 4k_2\cdot k_3 k_4\cdot k_5}=-{1\over s_{23}s_{45} },\label{R1}
\ee
where we use the notation $s_{ab}=(k_a+k_b)^2, s_{abc}=(k_a+k_b+k_c)^2$, etc. For the term $\R(w,f_4)$, we note that if $w=\delta$, $f_3=0$ implies $z=\O(\delta)$. Setting $w=xz$, 
\begin{alignat}{2}
f_3&=-zg_3+\O(\delta^2),\quad g_3&=k_3\cdot k_4+k_3\cdot k_5(1-x);\\
f_4&=-zg_4+\O(\delta^2),\quad g_4&=k_4\cdot k_3x+k_4\cdot k_5(x-1).\
\end{alignat}
Then
\be
\R(w,f_4)=-{1\over 4}\oint_{\O_{zg_4}}{z dzzdx\over (1-z)xz^3 g_3g_4}=-{1\over 4}\oint_{g_4=0}{dx\over x g_3g_4}
=-{1\over s_{345}}\left[{1\over s_{45}}+{1\over s_{34}}\right].\label{R2}
\ee
Noting that the integrand of (\ref{ampN5}) is symmetric under the interchanges $z\leftrightarrow 1- w, k_2\leftrightarrow k_5, k_3\leftrightarrow k_4, f_3\leftrightarrow f_4$, we deduce that 
\be
\R(f_3,1-z)=-\left[{1\over s_{23}}+{1\over s_{34}}\right]{1\over s_{234}}.\label{R3}
\ee
Putting together (\ref{R1}), (\ref{R2}) and (\ref{R3}), we see that $\A_5^\phi$ gives the sum of planar tree $\phi^3$ diagrams for momentum ordering $(k_1,k_2,k_3,k_4,k_5)$,
\be
\A_5^\phi={1\over s_{23}s_{234} }+{1\over s_{34}s_{234} }+{1\over s_{23}s_{45} }+{1\over s_{345}s_{34} }+{1\over s_{345}s_{45} }.\label{R4}
\ee
 This formula does not explicitly involve $k_1$ and hold when $k_1$ is off-shell, {\it i.e.} when $k_1^2\ne 0$, and the same is true of the form (\ref{amp4}) for $\A^\phi_4$.

\section{\bf  Inductive Proof for (partly) Off-Shell Massless ${\boldsymbol\phi}^{\bold 3}$ Theory}
\label{phi32}

We shall prove by induction that (\ref{ampphi2}) provides a formula for the sum of planar tree diagrams that is valid when $k_1$ is off-shell. We shall do this by establishing the recurrence relation
\begin{align}
A_N^\phi(k_1,k_2,\ldots,k_N)={1\over \bar s_3}&A_{N-1}^\phi(k_1+k_2,\ldots,k_N)\cr
&+\sum_{m=3}^{N-2}A_{m}^\phi(\pi_m,k_2,\ldots,k_m){1\over s_m\bar s_{m+1}}A_{N-m+1}^\phi(\bar\pi_{m+1},k_{m+1},\ldots,k_N)\cr
&\qquad+A_{N-1}^\phi(k_N+k_1,k_2,\ldots,k_{N-1}){1\over s_{N-1}}, \label{RR}
\end{align}
where
\be\pi_m=-k_2-\ldots-k_m,\quad \bar\pi_m=-k_m-\ldots-k_N, \quad s_m= \pi_m^2, \quad \bar s_m= \bar\pi_m^2,\label{defpi}\ee
which holds if $A_N^\phi$ denotes the sum of planar tree diagrams with the first momentum off-shell.  This relation follows from the observation that, for $N>3$, the external line carrying the off-shell momentum $k_1$ enters the tree diagram either at a vertex in which one of the other two lines meeting there is internal or both are. In the former case, because the diagrams are planar with the given ordering of momenta, the other external line must either carry momentum $k_2$ or momentum $k_N$, and the internal line is a propagator carrying off-shell momentum $k_1+k_2$ or $k_1+ k_N$ joining it to an $(N-1)$-point tree amplitude in which all the other momenta are the remaining $k_a$ in order, and so on shell, providing the first and last terms in (\ref{RR}). In the latter case, the two off-shell internal lines have momenta $\pi_m$ and $\bar\pi_{m+1}$. 

To establish (\ref{RR}) holds for $A_N^\phi=\A_N^\phi$ defined by (\ref{ampphi2}), we use the global residue theorem as we did in case $N=5$, again displacing the factors of  $1-z_3$ and $z_{N-1}$ by $\epsilon$ and $\delta$, respectively, in order to avoid singular configurations
\be
\A_N^\phi={\varepsilon_N\over 2^{N-3}}\oint_\O{1\over (1-\epsilon-z_3)(z_{N-1}-\delta)}\prod_{a=3}^{N-2}z_a\prod_{a=4}^{N-1}(1-z_a)\prod_{b=5}^{N-1}\prod_{a=3}^{b-2}(z_a-z_b)^2\prod_{a=3}^{N-1}{dz_a\over f_a} .\label{ampphi3}
\ee
Denoting the right hand side of (\ref{ampphi3}) by $\R(f_3,f_4,\ldots,f_{N-2},f_{N-1})$, using the global residue theorem,
\begin{align}
\R(f_3,f_4,&\ldots,f_{N-2},f_{N-1})=-\R(z_{N-1}^\delta,f_4,\ldots,f_{N-2},f_{N-1})\cr
&-\R(f_3,f_4,\ldots,f_{N-2},1-z_3^\epsilon)-\R(z_{N-1}^\delta,f_4,\ldots,f_{N-2},1-z_3^\epsilon),\label{res1}
\end{align}
where $z_3^\epsilon=z_3+\epsilon$ and $z_{N-1}^\delta=z_{N-1}-\delta$. Applying the global residue theorem again,
\begin{align}
\R(z_{N-1}^\delta,f_4,f_5,\ldots,f_{N-3},f_{N-2},1-z_3^\epsilon)=&-\R(z_{N-1}^\delta,f_{N-1},f_5,\ldots,f_{N-3},f_{N-2},1-z_3^\epsilon)\cr
-\R(z_{N-1}^\delta,f_{4},f_5,\ldots,f_{N-3},&f_{3},1-z_3^\epsilon)-\R(z_{N-1}^\delta,f_{N-1},f_5,\ldots,f_{N-3},f_{3},1-z_3^\epsilon),\cr
=(-1)^{N+1}[\R(z_{N-1}^\delta,f_5,\ldots,f_{N-3},f_{N-2},&f_{N-1},1-z_3^\epsilon)+\R(z_{N-1}^\delta,f_{3},f_{4},f_5,\ldots,f_{N-3},1-z_3^\epsilon)]\cr
&+\R(z_{N-1}^\delta,f_{3},f_5,\ldots,f_{N-3},f_{N-1},1-z_3^\epsilon).\label{res2}
\end{align}
Repeating this process, we obtain
\begin{align}
-\R(z_{N-1}^\delta&,f_4,f_5,\ldots,f_{N-3},f_{N-2},1-z_3^\epsilon)\cr
&=(-1)^{N}\sum_{m=3}^{N-2}\R(z_{N-1}^\delta,f_{3},f_4,\ldots,f_{m-1},f_{m+2},\ldots ,f_{N-2},f_{N-1},1-z_3^\epsilon)\cr
&=\sum_{m=3}^{N-2}\R(f_{3},f_4,\ldots,f_{m-1},1-z_3^\epsilon,z_{N-1}^\delta,f_{m+2},\ldots,f_{N-2},f_{N-1}).
\end{align}
After substituting this form for $\R(z_{N-1}^\delta,f_4,\ldots,f_{N-2},1-z_3^\epsilon)$  in (\ref{res1}), we shall show that the various terms correspond to the terms in the recurrence relation  (\ref{RR}).

First consider $\R(z_{N-1}^\delta,f_4,\ldots,f_{N-2},f_{N-1})$ as $\delta\rightarrow 0$. As $z_{N-1}=\delta\rightarrow 0$ with $f_{a}=0, 4\leq a \leq N-1$, for general values of the external momenta satisfying the zero-mass condition, we must  have  that $z_3$ also vanishes, for otherwise we would have $k_S^2=0$, as  in (\ref{grs}) with $s=N$, where $S$ is the set of $a$ for which $z_a$ vanishes as $\delta\rightarrow 0$  (including $N$), and we  have stipulated that
the $k_a$ are in general position. In fact the residue comes from the region where all $z_a=\O(\delta), 3\leq a\leq N-1$. We shall first calculate the contribution from this region and then show it is necessary for all $z_a, 3\leq a\leq N-1,$ to vanish to get a non-zero contribution. 

Set $z_a=x_az_3, 3\leq a\leq N,$ so that $x_3=1, x_N=0$. Then, as in (\ref{SES1}),
 for $3\leq a\leq N-1,$
\be
f_a(z)=-z_3^{N-4}g_a(x)[1+\O(\delta)], \label{fapprox}\ee
where
\be g_a(x)=\hat g_a(x)\prod_{c=3\atop c\ne a}^N(x_a-x_c)=\sum_{b=3\atop b\ne a}^Nk_a\cdot k_b\prod_{c=3\atop c\ne a,b}^N(x_a-x_c).
\ee
Then $\R(z_{N-1}^\delta,f_4,\ldots,f_{N-2},f_{N-1})$ is the residue of 
\begin{align}
-{\varepsilon_{N-1}\over 2^{N-3}}\oint{dz_3\over z_3x_{N-1}g_3}&\prod_{a=4}^{N-2}x_a\prod_{b=5}^{N-1}\prod_{a=3}^{b-2}(x_a-x_b)^2\prod_{a=4}^{N-1}{dx_a\over g_a}\cr
&=-{\varepsilon_{N-1}\over 2^{N-3}}\oint{1\over x_{N-1}g_3}\prod_{a=4}^{N-2}x_a\prod_{b=5}^{N-1}\prod_{a=3}^{b-2}(x_a-x_b)^2\prod_{a=4}^{N-1}{dx_a\over g_a}\label{zpole}
\end{align}
at $g_4=\ldots=g_{N-1}=0$.  In (\ref{zpole}), in the limit $\delta\rightarrow 0$, a factor of $\delta^{(N-4)^2}$ in the numerator, coming from the products
\be
\prod_{a=3}^{N-2}z_a\prod_{b=5}^{N-1}\prod_{a=3}^{b-2}(z_a-z_b)^2\label{numprod}
\ee
in (\ref{ampphi3}), is cancelled by an equal factor in the denominator, coming from 
\be
\prod_{a=3}^{N-1}{dz_a\over f_a},\label{denomprod}
\ee
leaving a finite answer.  As in (\ref{SES2}), 
\be \sum_{a=3}^{N-1}x_a\hat g_a={\bar s_3\over 2},\label{gsum}\ee
So, when $g_4=\ldots=g_{N-1}=0$,
\be g_3=\hat g_3\prod_{c=4}^N(x_3-x_c)={\bar s_3\over 2}\prod_{c=4}^{N-1}(1-x_c),\ee
and, provided that there are no other contributions to this residue from other regions,  $\R(z_{N-1}^\delta,f_4,\ldots,f_{N-2},f_{N-1})$ equals
\begin{align}
-{\varepsilon_{N-1}\over 2^{N-4}\bar s_3}\oint{1\over x_{N-1}(1-x_4)}\prod_{a=4}^{N-2}x_a&\prod_{a=5}^{N-1}(1-x_a)\prod_{b=6}^{N-1}\prod_{a=4}^{b-2}(x_a-x_b)^2\prod_{a=4}^{N-1}{dx_a\over g_a}\cr
&=-{1\over \bar s_3} A^\phi_{N-1}(k_1+k_2,k_3,k_4,\ldots, k_{N-1},k_N),\label{A1}
\end{align}
by induction.

If we consider contributions to $\R(z_{N-1}^\delta,f_4,\ldots,f_{N-2},f_{N-1})$ from regions in which not all the $z_a, 3\leq a\leq N-1$ vanish as $\delta\rightarrow 0$. As we have argued, $z_3\rightarrow 0$ as well as $z_{N-1}$ as $\delta\rightarrow 0$. If the set $S=\{a:z_a\rightarrow 0\}$ contains only $M$ of the $z_a, 3\leq a\leq N$, where $M<N-2$, the product (\ref{denomprod}) in  (\ref{ampphi3}) will contribute a factor of $\delta^{(M-2)^2}$ to the denominator, but the product (\ref{numprod}) will contribute a factor of $\delta^L$ where $L>(M-2)^2$, which implies that the contribution vanishes as $\delta\rightarrow 0$. This is because the products in (\ref{numprod}) would contribute an exactly balancing contribution of $\delta^{(M-2)^2}$ if  the $a\in S$ were consecutive but the second product in  (\ref{numprod})  contributes a factor of $\delta^2$ in the limit for every $a,b\in S$ with $3\leq a<b\leq N-1$ and $b-a>1$, and this number is at a minimum for a given $M$ when the $a$ are consecutive. However, as $3, N-1\in S$ and $M<N-2$, the $a\in S$ cannot be consecutive and so the contribution vanishes in the limit, establishing (\ref{A1}).

Because of the symmetry under $z_a\leftrightarrow 1-z_{N+2-a}, k_a\leftrightarrow k_{N+2-a}, 2\leq a\leq N,$ we have the corresponding result
\be
\R(f_3,f_4,\ldots,f_{N-2},1-z_3^\epsilon)=-A_{N-1}^\phi(k_N+k_1,k_2,\ldots,k_{N-1}){1\over s_{N-1}}.\label{A2}
\ee

It remains to consider $\R(f_{3},f_4,\ldots,f_{m-1},1-z_3^\epsilon,z_{N-1}^\delta,f_{m+2},\ldots,f_{N-2},f_{N-1})$. 
This contribution  comes from the region 
$ z_a=\O(\delta), \; m+1\leq a\leq N-1;\; 1-z_a=\O(\epsilon),\; 3\leq a\leq m, $ for reasons similar to those that determined the appropriate regions for 
the residues already considered: if $S_0$ denotes the set of $a$ for which $z_a\rightarrow 0$ (including $a=N$) and $S_1$ denotes the set of $a$ for which $z_a\rightarrow 1$ (including $a=2$), then each of $S_0,S_1$ much 
contain one of $m$ or $m+1$ or we should have $k_{S_0}^2=0$ or $k_{S_1}^2=0$, contradicting the assumption that the $k_a$ are in general position. If $S_0$ does not consist of consecutive elements we should find again that the contribution will vanish in the limit and similarly for $S_1$. Thus the only way to get a non-zero contribution is to have $S_0=\{a: m+1\leq a\leq N\}$ and $S_1=\{a:2\leq a\leq m\}$, which is the region we are considering.

Set $ z_a=x_az_{m+1},\; m+1\leq a\leq N;\; 1-z_a=y_a(1-z_{m}),\;2\leq a\leq m$, 
so that $x_{m+1}=1, x_N=0$ and $y_{m}=1, y_{2}=0$.  Then, as in (\ref{SES1}), for $3\leq a\leq m$,
\be
f_a(z)=(-1)^{m+1}(1-z_{m})^{m-3}h_a(y)[1+\O(\delta,\epsilon)],\ee
where
\be h_a(y)=\hat h_a(y)\prod_{c=2\atop c\ne a}^{m}(y_a-y_c).=\sum_{b=2\atop b\ne a}^{m}k_a\cdot k_b\prod_{c=2\atop c\ne a,b}^{m}(y_a-y_c),
\ee
and, for $m+1\leq a\leq N-1$,
\be
f_a(z)=(-1)^{m+1}z_{m+1}^{N-m-2}g_a(x)[1+\O(\delta,\epsilon)],\ee
where
\be g_a(x)=\hat g_a(x)\prod_{c=m+1\atop c\ne a}^N(x_a-x_c)=\sum_{b=m+1\atop b\ne a}^Nk_a\cdot k_b\prod_{c=m+1\atop c\ne a,b}^N(x_a-x_c).\ee

Then $\R(f_{3},f_4,\ldots,f_{m-1},1-z_3^\epsilon,z_{N-1}^\delta,f_{m+2},\ldots,f_{N-2},f_{N-1})$ is the residue at $g_{m+2}=\ldots=g_{N-1}=0$ and $h_{3}=\ldots=h_{m-1}=0$ of 
\begin{align}
{\varepsilon_N\varepsilon_{N,m} \over 2^{N-3}}\int&\prod_{a=3}^{m-1}{dy_a\over h_a}\prod_{a=3}^{m-2}(1-y_a)^2\prod_{a=4}^{m-1}y_a\prod_{a=3}^{m-3}\prod_{b=a+2}^{m-1}(y_a-y_b)^2\times {dz_{m}dz_{m+1}\over  (1-z_3)z_{N-1}h_{m}g_{m+1}}\cr
&\qquad\qquad\times\prod_{a=m+2}^{N-1}{dx_a\over g_a}\prod_{a=m+2}^{N-2}x_a\prod_{b=m+3}^{N-1}(1-x_b)^2\prod_{b=m+4}^{N-1}\prod_{a=m+2}^{b-2}(x_a-x_b)^2,\label{res3}
\end{align}
where $\varepsilon_{N,m} =(-1)^{(m+1)N}$. At $g_{m+2}=\ldots=g_{N-1}=0$, and $h_{3}=\ldots=h_{m-1}=0$, as in (\ref{SES2}),
\be g_{m+1}=\bar s_{m+1}\prod_{c=m+2}^{N-1}(1-x_c),\quad h_{m}=s_{m}\prod_{c=3}^{m-1}(1-y_c),\ee
so that (\ref{res3}) becomes
\begin{align}
  &-\int{\varepsilon_{m}\over 2^{m-3}(1-y_3)y_{m-1}}\prod_{a=3}^{m-1}{dy_a\over h_a}\prod_{a=3}^{m-2}(1-y_a)\prod_{a=4}^{m-1}y_a\prod_{a=3}^{m-3}\prod_{b=a+2}^{m-1}(y_a-y_b)^2{1\over s_{m}}\times
  {dz_{m}dz_{m+1}\over (1-z_{m})z_{m+1}}\cr
&\times{\varepsilon_{N-m+1}\over \bar s_{m+1}2^{N-m-2}(1-x_{m+2})x_{N-1}}\prod_{a=m+2}^{N-1}{dx_a\over g_a}\prod_{a=m+2}^{N-2}x_a\prod_{b=m+3}^{N-1}(1-x_b)\prod_{b=m+4}^{N-1}\prod_{a=m+2}^{b-2}(x_a-x_b)^2,
\end{align}
giving, on integration,
$$A_{m}^\phi(\pi_m,k_2,\ldots,k_m){1\over s_m\bar s_{m+1}}A^\phi_{N-m+1}(\bar\pi_{m+1},k_{m+1},\ldots,k_N)$$
as a contribution to the amplitude $A_N^\phi$, and, together with (\ref{A1}) and (\ref{A2}), establishing the recurrence relation (\ref{RR}).

\section{\bf  BCFW Proof for Massless ${\boldsymbol\phi}^{\bold 3}$}
\label{phi33}

The proof given in section \ref{phi32} that, with $\widehat\Psi_N$ constant, we obtain the tree amplitudes of massless $\phi^3$, would be difficult to extend to the gauge theory case, so, as a preliminary to discussing this case in section \ref{GT}, we now give a proof using the BCFW relations. To this end, choose a (possibly complex) momentum vector $\ell$ such that $\ell^2=\ell\cdot k_2=\ell\cdot k_N=0,$ and consider
\be
A^\phi_N(\zeta)=A_N^\phi(k_1,k_2+\zeta\ell,k_3,\ldots,k_{N-1},k_N-\zeta\ell),\label{Azeta}
\ee
where $A^\phi_N$ denotes the sum of planar tree diagrams. Then $A^\phi_N(\zeta)$ is meromorphic in $\zeta$ with poles arising because of the propagators in the tree diagrams it comprises. The residues of these poles are given by the product of amplitudes $A^\phi_m$, where $m<N$ and so $A^\phi_N(\zeta)$ can be expressed as a sum over these residues, and this provides a recursion relation, which is the BCFW relation \cite{BCFW,AK}. 

If it were not the case that $A^\phi_N(\zeta)\rightarrow 0$ as $\zeta\rightarrow \infty$, a constant would have to be included, in essence a boundary term from infinity. This would indeed arise if we were considering the sum of all $N$-point tree functions regardless of the order of the external momenta, but, the ordering of momenta in $A^\phi_N$ and the fact that the displaced momenta $k_2+\zeta\ell$ and $k_N-\zeta\ell$ are not adjacent means that at least one propagator denominator in each tree diagram included in $A_N^\phi$ depends linearly on $\zeta$ and so $A^\phi_N(\zeta)\rightarrow 0$ as $\zeta\rightarrow \infty$.

Integrating around a small contour $\O_\zeta$ enclosing $\xi=\zeta$ and then moving the contour out to infinity,
\be A_N^\phi(\zeta)=\oint_{\O_\zeta}{A_N^\phi(\xi)d\xi\over \xi-\zeta}=-\sum_{i}{\hbox{Res}_{\zeta_i}A_N^\phi\over \zeta_i-\zeta},\label{CL}\ee
where the poles of $A_N^\phi(\zeta)$ occur at the points $\zeta=\zeta_i$. Define $\pi_m,\bar\pi_m, s_m, \bar s_m$ as in (\ref{defpi}), and $\pi_m^\zeta$ and $\bar\pi_m^\zeta$ by replacing $k_2$ by $k_2^\zeta$ in $\pi_m$ and $k_N$ by $k_N^\zeta$ in $\bar\pi_m$, respectively, where 
\be
k_2^\zeta=k_2+\zeta\ell,\qquad  k_N^\zeta=k_N-\zeta\ell. \label{defkzeta}
\ee
Then the poles of $A^\phi_N(\zeta)$ occur where $(\pi_m^\zeta)^2=0$ or $(\bar\pi_m^\zeta)^2=0$,  {\it i.e.} at
\be\zeta=s_m/2\pi_m\cdot\ell\equiv \zeta_m^L,\quad\hbox{and}\quad \zeta=-\bar s_m/2\bar\pi_m\cdot\ell\equiv \zeta_m^R,\qquad 3\leq m\leq N-1,\label{zetaLR}\ee
with residues given by
\be\hbox{Res}_{\zeta^L_m}A_N^\phi=-A^\phi_m(\pi_m^{\zeta^L_m}, k_2^{\zeta^L_m},k_3,\ldots,k_m){1\over 2\pi_m\cdot\ell}A^\phi_{N-m+2}(k_1,-\pi_{m}^{\zeta^L_m},k_{m+1},\ldots,k_{N-1},k_N^{\zeta^L_m}),\label{ResL}
\ee
and
\be\hbox{Res}_{\zeta^R_m}A_N^\phi=A^\phi_m(k_1,k_2^{\zeta^R_m},k_3,\ldots,k_{m-1},-\bar\pi_{m}^{\zeta^R_m} ){1\over 2\bar\pi_m\cdot\ell}A^\phi_{N-m+2}(\bar\pi_{m}^{\zeta^R_m},k_{m},\ldots,k_{N-1},k_N^{\zeta^R_m}).\label{ResR}\ee
Then, putting $\zeta=0$ in (\ref{CL}),
\be A_N^\phi(k_1,k_2,\ldots,k_N)=A^\phi_N(0)=-\sum_{m=3}^{N-1}\left[{\pi_m\cdot\ell\over s_m}\hbox{Res}_{\zeta^L_m}A_N^\phi-{\bar\pi_m\cdot\ell\over \bar s_m}\hbox{Res}_{\zeta^R_m}A_N^\phi\right].\label{RR1}\ee
This relation is sufficient to determine $A_N^\phi$ for $N>3$ from $A_3^\phi=1$. We now seek to show $A_N^\phi=\A_N^\phi$, as defined by (\ref{ampphi2}) satisfies (\ref{RR1}), with $\hbox{Res}_{\zeta^L_m}A_N^\phi$ and $\hbox{Res}_{\zeta^R_m}A_N^\phi$ given by (\ref{ResL}) and (\ref{ResR}), respectively.

Defining  $\A^\phi_N(\zeta)$ analogously to $A^\phi_N(\zeta)$,
\be\A^\phi_N(\zeta)={\varepsilon_N\over 2^{N-3}}\oint{1\over (1-z_3)z_{N-1}}\prod_{a=3}^{N-2}z_a\prod_{a=4}^{N-1}(1-z_a)\prod_{b=5}^{N-1}\prod_{a=3}^{b-2}(z_a-z_b)^2\prod_{a=3}^{N-1}{dz_a\over f_a(z,\zeta)}\label{Aphizeta}
\ee
where $f_a(z,\zeta)$ is obtained from $f_a(z)$ by $k_2\mapsto k_2^\zeta, k_N\mapsto k_N^\zeta$,
\be
f_a(z,\zeta)=\prod_{c=2\atop c\ne a}^N(z_a-z_c)\left[{k_a\cdot (k_2+\zeta\ell)\over (z_a-1)}+\sum_{b=3\atop b\ne a}^{N-1}{k_a\cdot k_b\over (z_a-z_b)}+{k_a\cdot (k_{N}-\zeta\ell)\over z_a}\right]
\ee
and the contour encloses the solutions of 
\be
f_a(z,\zeta)=0,\qquad 3\leq a\leq N-1,\label{f3N1}
\ee
 for given $\zeta$, but excludes the poles at $z_{N-1}=0$ and $z_3=1$. 

The poles of $\A^\phi_N(\zeta)$ occur when the contour is pinched between the zeros of the $f_a$ and either $z_{N-1}=0$ or  $z_3=1$. First consider a solution of (\ref{f3N1}) which is such that, as $\zeta\rightarrow \zeta_0$. $z_{N-1}\rightarrow 0$. It may be that as $\zeta\rightarrow\zeta_0$ some other $z_a\rightarrow0$. Suppose again that $S$ denotes the set of $a$, with $3\leq a\leq N$, for which $z_a\rightarrow 0$. For essentially the same reasons used in section \ref{phi32} to determine which contributions to the recurrence relation considered there were nonzero, having $z_a\rightarrow 0$ for $a\in S$ will not produce a pole unless the $a$ in $S$ are consecutive. Thus we must have $S=\{a: m\leq a\leq N\}$ for some $m$ with $3\leq m\leq N-1$. Then, from (\ref{grs}), $k_S^2=(\bar\pi_m^{\zeta_0})^2=0$, so that $\zeta_0=\zeta_m^R$, as defined by (\ref{zetaLR}), and the solution to (\ref{f3N1}) has $z_a\rightarrow 0$,  $m\leq a\leq N-1$, as  $\zeta\rightarrow\zeta_m^R$. Similarly, there are solutions to (\ref{f3N1}) that have the property that $z_a\rightarrow 1$, $3\leq a\leq m$, as  $\zeta\rightarrow\zeta_m^L$, as defined by (\ref{zetaLR}).

We now investigate the (potential) pole at $\zeta=\zeta_m^R$, which  comes from the region of integration in (\ref{Aphizeta}) given by 
\be z_a=\O(\delta), \qquad m\leq a\leq N-1.\label{regpole}\ee
Set
\be z_a=x_az_{m},\; m\leq a\leq N;\qquad\hbox{so}\qquad x_{m}=1, x_N=0.\label{zxz}\ee

As in (\ref{SES1}) and (\ref{fapprox}), for $m\leq a\leq N-1$,
\be
f_a(z,\zeta)=(-1)^{m}z_{m}^{N-m-1}g_a(x,\zeta)\prod_{c=2}^{m-1}z_c\;[1+ \O(\delta) ],
\ee
where
\be
g_a(x,\zeta)= \hat g_a(x,\zeta)\prod_{c=m\atop c\ne a}^N(x_a-x_c)=\sum_{b=m\atop b\ne a}^Nk_a\cdot k_b^\zeta\prod_{c=m\atop c\ne a,b}^N(x_a-x_c).\ee
As in (\ref{SES2}) and (\ref{gsum}),
\be
\sum_{a=m}^{N-1}x_a\hat g_a(x,\zeta)={1\over 2}\left(\sum_{a=m}^{N}k_a^\zeta\right)^2=(\zeta-\zeta_m^R)\ell\cdot\bar\pi_m
\ee
So when $g_a(z,\zeta)=0, \quad m+1\leq a\leq N-1$, 
\be
g_m(x,\zeta)=(\zeta-\zeta_m^R)\ell\cdot\bar\pi_m\prod_{c=m+1}^{N-1}(1-x_c).
\ee
For $3\leq a\leq m-1$,
\be
f_a(z,\zeta)=z_{a}^{N-m}h_a(z,\zeta)[1+ \O(\delta) ],
\ee
where 
\be
 h_a(z,\zeta)=\sum_{b=2\atop b\ne a}^{m-1}k_a\cdot k_b^\zeta z_a\prod_{c=2\atop c\ne a,b}^{m-1}(z_a-z_c)
- k_a\cdot \bar\pi^\zeta_m\prod_{c=2\atop c\ne a}^{m-1}(z_a-z_c).
\ee
Noting $\varepsilon_N(-1)^{m(N+1)}=-\varepsilon_m\varepsilon_{N-m+2}$
\begin{align}
\hbox{Res}_{\zeta^R_m}&\A^\phi_N(\zeta)=\oint{d\zeta\over 2\ell\cdot\bar\pi_m(\zeta-\zeta_m^R)}\cr
&\times{\varepsilon_m\over 2^{m-3}(1-z_3)z_{m-1}} \prod_{a=3}^{m-2}z_a\prod_{a=4}^{m-1}(1-z_a)\prod_{b=5}^{m-1}\prod_{a=3}^{b-2}(z_a-z_b)^2\prod_{a=3}^{m-1}{dz_a\over h_a}
\times \left[-{dz_{m}\over z_{m}}\right] \cr
&\times{\varepsilon_{N-m+2}\over 2^{N-m-1}(1-x_{m+1})x_{N-1}} \prod_{a=m+1}^{N-2}x_a\prod_{b=m+2}^{N-1}(1-x_b)
\prod_{b=m+3}^{N-1}\prod_{a=m+1}^{b-2}(x_a-x_b)^2\prod_{a=m+1}^{N-1}{dx_a\over g_a}\cr
\end{align}

Integrating with respect to $\zeta$ and $z_m$ shows that $\hbox{Res}_{\zeta^R_m}\A_N^\phi$ provides the appropriate contribution to the BCFW relation,
\be\hbox{Res}_{\zeta^R_m}\A_N^\phi=\A^\phi_m(k_1,k_2^{\zeta^R_m},k_3,\ldots,k_{m-1},-\bar\pi_{m}^{\zeta^R_m}){1\over 2\bar\pi_m\cdot\ell}A^\phi_{N-m+2}(\bar\pi_{m}^{\zeta^R_m},k_{m},\ldots,k_{N-1},k_N^{\zeta^R_m}),\ee
since the pole at $z_m=0$ is exterior to the contour. 
The residue at $\zeta=\zeta_m^L$ follows similarly, using the symmetry under $m\leftrightarrow N+2 -m, z_a\leftrightarrow 1-z_a, L \leftrightarrow R,$ establishing the full BCFW relation and so proving the formula for $A^\phi_N$ by induction.

\section{\bf  Proof for Pure Gauge Theory}
\label{GT}

The CHY proposal \cite{CHY1, CHY2} for a choice of $\widehat\Psi_N(z;k;\epsilon)$ that gives the tree amplitudes for pure gauge theory in any space-time dimension employs the Pfaffian of the antisymmetric matrix 

\be
\Psi_N(z;k;\epsilon)= \left(\begin{matrix}  A&D\cr C&B\cr\end{matrix} \right),
\ee
where 
\be
A_{ab}={k_{a}\cdot k_{b}\over z_{a}- z_{b}},\quad B_{ab}= {\epsilon_{a}\cdot \epsilon_{b}\over z_{a}- z_{b}},\quad C_{ab}= {\epsilon_{a}\cdot k_{b}\over z_{a}- z_{b}},\; a\ne b, \quad 1\leq a,b\leq N;
\ee
\be
A_{aa}=B_{aa}=0,\quad C_{aa}=-\Sigma_a,\quad \Sigma_a=\sum_{c=1\atop c\ne a}^N{\epsilon_{a}\cdot k_{c}\over z_{a}- z_{c}},\quad1\leq a\leq N,
\ee
$D_{ab}=-C_{ba},\; 1\leq a,b\leq N$. [Properties of Pfaffians are reviewed in appendix \ref{Pfaff}, together with further some results needed here for our discussion.]

The Pfaffian of $\Psi_N(z;k;\epsilon)$ itself is zero when the conditions (\ref{SE}) hold, because then it has two null vectors arising from common null vectors of $A$ and $C$: the vectors $(1,1,\ldots,1)$ and 
$(z_1,z_2,\ldots,z_N)$. So, CHY consider the Pfaffian of the $(2N-2)\times(2N-2)$-dimensional matrix $\Psi_N^{(a,b)}(z;k;\epsilon)$ obtained by omitting the $a$-th and $b$-th rows and columns of $\Psi_N(z;k;\epsilon)$ and
demonstrate that 
\be
\Pf'\Psi_N(z;k;\epsilon)=2{(-1)^{a+b}\over z_a-z_b}\Pf\Psi_N^{(a,b)}(z;k;\epsilon)
\ee
is independent of $a,b$  for $1\leq a<b\leq N$. Then they give evidence for the proposition that the tree amplitudes, $\A^\YM_N$, for gauge theory are given by taking 
\be
\widehat\Psi_N(z;k;\epsilon)=-(-2)^{3-N}\Pf'\Psi_N(z;k;\epsilon)\prod_{a=1}^N(z_a-z_{a+1})
\ee
in (\ref{amp}), where the normalization factor has been chosen for the convenience of the present discussion. The required M\"obius invariance of $\widehat\Psi_N(z;k;\epsilon)$ is straightforward to verify.

As in (\ref{ampphi2}),
\be
\A^\YM_N(k;\epsilon)={\varepsilon_N\over 2^{N-3}}\oint_\O{\widetilde\Psi_N^o(z;k;\epsilon)\over (1-z_3)z_{N-1}}\prod_{a=3}^{N-2}z_a\prod_{a=4}^{N-1}(1-z_a)\prod_{b=5}^{N-1}\prod_{a=3}^{b-2}(z_a-z_b)^2\prod_{a=3}^{N-1}{dz_a\over f_a}, \label{AYM}
\ee
where again $\varepsilon_N=(-1)^{\half N(N+1)}$, $f_a$ is given by (\ref{fa}), the contour $\O$ encircles the simultaneous solutions of $f_a(z,k)=0, \;3\leq a\leq N-1$ (but excludes $z_3=1$ and $z_{N-1}=0$), and where
\be
\widetilde\Psi_N^o(z;k;\epsilon)
=\lim_{z_1\rightarrow\infty}\left(z_1^2\Pf'\Psi_N(z;k;\epsilon)\right)\prod_{a=2}^{N-1}(z_a-z_{a+1}).\label{PsiN1}
\ee
Now,
\be
\lim_{z_1\rightarrow\infty}z_1^2\Pf'\Psi_N(z;k;\epsilon)=\Pf'\Psi^o_N(z;k;\epsilon),
\ee
where $\Psi^o_N(z;k;\epsilon)\equiv \Psi^o$ is antisymmetric and 
\be
\Psi^o_{ab}(z)=\Psi_{ab}(z), \quad \hbox{if}\quad a,b\ne 1,N+1, \label{Psio}
\ee
\be
\Psi^o_{1b}={k_{1}\cdot k_{b}},\; \Psi^o_{N+1,N+b}=\epsilon_{1}\cdot \epsilon_{b},\;1\leq b\leq N;
\ee
\be
\Psi^o_{1,N+1}=\sum_{c=2}^Nz_c\epsilon_{1}\cdot k_{c};\quad\Psi^o_{1,N+b}= k_{1}\cdot \epsilon_{b},\;2\leq b\leq N;\quad
\Psi^o_{a,N+1}= -\epsilon_{1}\cdot k_a,\;2\leq a\leq N;
\ee
and 
\be
\Pf'\Psi^o={(-1)^{a+b}\over z_a-z_b}\Pf \Psi^o_{(ab)}=(-1)^{1+a}\Pf \Psi^o_{(1a)}
\ee
 independent of $a,b$  for $1< a<b\leq N$, $\Psi^o_{(ab)}$ being defined relative to $\Psi^o$ in the same way that $\Psi_{(ab)}$ is relative to $\Psi$. Thus, in (\ref{AYM}),  (\ref{PsiN1}),
 \be
 \widetilde\Psi_N^o(z;k;\epsilon)=(-1)^N\Pf \Psi_N^o(z;k;\epsilon)_{(2N)}\prod_{a=1}^N(z_a-z_{a+1}).\label{PsiN2}
 \ee
 
 We now give a proof that $\A^\YM_N$, defined by (\ref{AYM}), does indeed give the $N$-gluon Yang-Mills tree amplitude by establishing that it satisfies a BCFW recurrence relation, as in (\ref{ResL}-\ref{RR1}). To this end, we need to consider $\A^\YM_N(\zeta)$, obtained from $\A^\YM_N$ by deforming the momenta $k_2\rightarrow k_2^\zeta=k_2+\zeta \ell,\; k_N\rightarrow k_N^\zeta=k_N-\zeta \ell, $ as in (\ref{defkzeta}),  where, again, $\ell^2=\ell\cdot k_2=\ell\cdot k_N=0$, and also by deforming the basis of polarization vectors $\epsilon_2^{\zeta 1}, \epsilon_2^{\zeta 2}, \epsilon_2^{ j},\; 3\leq j\leq d-2,$ for $k_2^\zeta$, as in \cite{AK}, so that
 \be
 \epsilon^{\zeta 1}_2+i \epsilon^{\zeta 2}_2=\bar\ell-2(\zeta/k_2\cdot k_N)k_N,\qquad \epsilon^{\zeta 1}_2-i \epsilon^{\zeta 2}_2=\ell
 \ee
where we have introduced $\bar\ell$ such that  $\bar\ell^2=\bar\ell\cdot k_2=\bar\ell\cdot k_N=0,\;\bar\ell\cdot\ell=2$, so that $\epsilon^{\zeta i}_2\cdot  \epsilon^{\zeta j}_2= \delta^{ij},\; \epsilon^{\zeta j}_2\cdot k_2^\zeta=0, \; 1\leq i,j\leq N-2,$ with
$\epsilon^{\zeta j}_2\equiv \epsilon^{ j}_2,\; 3\leq j\leq N-2,$ constant; and similarly for the basis of polarization vectors $\epsilon_N^{\zeta 1}, \epsilon_N^{\zeta 2}, \epsilon_N^{ j},\; 3\leq j\leq d-2,$ for $k_N^\zeta$.

So we now consider 
\be
\A^\YM_N(\zeta)=\A^\YM_N(k_1,k_2^\zeta,\ldots,k_{N-1},k_N^\zeta;\epsilon_1,\epsilon_2^\zeta,\ldots,\epsilon_{N-1},\epsilon_N^\zeta).
\ee
The difference between $\A^\YM_N(\zeta)$ here and $\A^\phi(\zeta)$ in section \ref{phi33} is the factor of $\widetilde\Psi_N^o(z;k;\epsilon)$, as defined by (\ref{PsiN2}), in the integrand of (\ref{AYM}), and this does not affect the nature of the singularities in $\zeta$,  which are thus still poles  at $\zeta=\zeta_m^L, \zeta=\zeta_m^R$, $3\leq m\leq N-1$, as defined in (\ref{zetaLR}). 
[The singularities of $\widetilde\Psi_N^o(z;k;\epsilon)$ occur when two or more of the $z_a$ coincide. When the $z_a$ are adjacent the calculation in appendix \ref{Pfaff} shows that factors of $z_a-z_{a+1}$ cancel those in the denominator, so that $\widetilde\Psi_N^o(z;k;\epsilon)$ is regular there. If they are not adjacent, these  are cancelled by the factors of $(z_a-z_b)^2, a>b+1,$ as in section \ref{phi32}.] To demonstrate that $\A^\YM_N$ satisfies the BCFW recurrence relation we need to show that $\widetilde\Psi_N^o$ factorizes appropriately at these values of $\zeta$.

The residue of pole of $\A^\phi(\zeta)$ at  $\zeta=\zeta_m^R$ comes from the region (\ref{regpole}). So again we set $z_a=x_a z,\; m\leq a\leq N,$ as in (\ref{zxz}), so that $x_{m}=1, x_N=0$. 
Then, as $z_m\rightarrow 0$, $\Psi^o_{ab}(z)$, as defined by (\ref{Psio}), is given by
\begin{alignat}{7}
\Psi^o_{ab}&\sim{k_a\cdot k_b\over z_a-z_b}, &\vphantom{{k_a\cdot k_b\over z_a-z_b}} &\;2\leq a<b< m;
&\;\Psi^o_{N+a,N+b}&\sim{\epsilon_a\cdot \epsilon_b\over z_a-z_b},
&\vphantom{{\epsilon_a\cdot \epsilon_b\over z_a-z_b}} &\;2\leq a<b< m;\cr
\Psi^o_{ab}&\sim{k_a\cdot k_b\over z_a}, &\vphantom{{k_a\cdot k_b\over z_a}} &\;2\leq a< m\leq b\leq N;
&\;\Psi^o_{N+a,N+b}&\sim{\epsilon_a\cdot \epsilon_b\over z_a},
&\vphantom{{\epsilon_a\cdot \epsilon_b\over z_a}} &\;2\leq a< m\leq b\leq N;\cr
\Psi^o_{ab}&\sim{k_a\cdot k_b\over z_m (x_a-x_b)},&\vphantom{{k_a\cdot k_b\over z_m (x_a-x_b)}} &\;m\leq a<b\leq N;
&\;\Psi^o_{N+a,N+b}&\sim{\epsilon_a\cdot \epsilon_b\over z_m (x_a-x_b)},
&\vphantom{{\epsilon_a\cdot \epsilon_b\over z_m (x_a-x_b)}} &\;m\leq a<b\leq N;\cr
\Psi^o_{a,N+b}&\sim{k_a\cdot \epsilon_b\over z_a-z_b},&\vphantom{{k_a\cdot \epsilon_b\over z_a-z_b}} &\;2\leq a,b< m,\;a\ne b;&\;\Psi^o_{a,N+b}&\sim-{k_a\cdot \epsilon_b\over z_b},
&\vphantom{-{k_a\cdot \epsilon_b\over z_b}} &\;2\leq b< m\leq a\leq N;\cr
\Psi^o_{a,N+b}&\sim{k_a\cdot \epsilon_b\over z_a}, &\vphantom{{k_a\cdot \epsilon_b\over z_a}} &\;2\leq a< m\leq b\leq N;
&\;\Psi^o_{a,N+b}&\sim{k_a\cdot \epsilon_b\over z_m (x_a-x_b)},
&\vphantom{{k_a\cdot \epsilon_b\over z_m (x_a-x_b)}} &\;m\leq a,b\leq N,\;a\ne b;\cr
\Psi^o_{a,N+a}&\sim\hat\Sigma_a, &\vphantom{\hat\Sigma_a} &\;2\leq a< m;
&\;\Psi^o_{a,N+a}&\sim{1\over z_m }\hat\Sigma_a,
&\vphantom{{1\over z_m }\hat\Sigma_a} &\;m\leq a\leq N.\label{zminfy}
\end{alignat}
where
\be
\hat\Sigma_1=\sum_{c=2}^{m-1}z_c\epsilon_{1}\cdot k_{c};\quad
   \hat\Sigma_a=\sum_{c=2\atop c\ne a}^{m-1}{\epsilon_{a}\cdot k_{c}\over z_{a}- z_{c}}+\sum_{c=m}^N{\epsilon_{a}\cdot k_{c}\over z_{a}}, \; 2\leq a<m;\ee
   \be
   \hat\Sigma_a=\sum_{c=m\atop c\ne a}^N{\epsilon_{a}\cdot k_{c}\over x_{a}- x_{c}}, \; m\leq a\leq N.\ee
   
Now rearrange the columns of $\Psi^o_{ab}$ by defining $u(a)=a$ for $1\leq a <m$ or $N+m\leq a\leq 2N$; $u(a)=N+a-m+1$ for $m\leq a \leq 2m-2$;
$u(a)=N+a-m+1$ for $2m-2< a \leq N+m-2$; and set 
\be
\Theta_{ab}=\Psi^o_{u(a)u(b)}.
\ee
so that $\Theta_{ab}=-\Theta_{ba}$. 
Then
\be
\Theta=\left[\begin{matrix}  E&F\cr -F^T&z_m^{-1}G\cr  \end{matrix} \right],\ee
where
\begin{align}
E&=\Psi^o_m(k_1,\ldots,k_{m-1},-\bar\pi_m;\epsilon_1,\ldots,\epsilon_{m-1},\epsilon^{s};z_3,\ldots,z_{m-1})_{(m,2m)}\cr
G&=\Psi^o_{N-m+2}(\bar\pi_m,k_m,\ldots,k_N;\epsilon^{s},\epsilon_m,\ldots,\epsilon_N;x_{m+1},\ldots,x_{N-1})_{(1,N-m+3)}
\end{align}
where $\bar\pi_m=-k_m-\ldots-k_N$, as in (\ref{defpi}), and $F_{ab}=\alpha_a\cdot\beta_b$, with
\be\alpha_1=k_1, \quad\alpha_a={k_a\over z_a}, \;2\leq a<m,\quad\alpha_m=\epsilon_1,\quad\alpha_a={\epsilon_a\over z_a},\;m<a\leq 2m-2,\ee
\be\beta_b=k_b,\;1\leq b\leq N-m+2, \quad \beta_b=\epsilon_b,\;N-m+3\leq b\leq 2N-2m+4.\ee
[Note that $\epsilon_s$, $\bar\pi_m$ do not appear directly in the expressions for $E, G$ because of the rows omitted in $[\Psi^o_m]_{(m,2m)}$ and
$[\Psi^o_{N-m+2}]_{(1,N-m+3)}$, respectively.]

Now, it is $\Psi^o_{(2,N)}$ that we want to consider rather than $\Psi^o$; replacing $\Psi^o$ with $\Psi^o_{(2,N)}$ effectively replaces $\Theta$ with 
\be\Theta_{(2,N+m-1)}=\left[\begin{matrix}  E_{(2)}&F_{[2,1]}\cr -F_{[2,1]}^T&z_m^{-1}G_{(1)}\cr  \end{matrix} \right],\ee
where $E_{(a)}$ indicates that the $a$-th row and column have been removed from $E$, and similarly for $G_{(a)}$, and $F_{[a,b]}$ indicates that the $a$-th row
 and $b$-th column have been removed from $F$. 
\be\Pf\Psi^o_{(2N)}=\Pf\Theta_{(2,N+m-1)}\ee
 So, using the Factorization Lemma of appendix \ref{Pfaff}, as $z_m\rightarrow 0$,
\be\Pf\Phi(z) \sim z_m^{-(N-m)}\sum_r\Pf \widetilde E^r\times \Pf \widetilde G^r,\label{asymPhi}\ee
 where
\be \widetilde E^r=\left[\begin{matrix}  E_{(2)}&(\alpha^r_{[2]})^T\cr -\alpha_{[2]}^r&0\cr  \end{matrix} \right],\qquad
  \widetilde G^r=\left[\begin{matrix}  0&(\beta_{[1]}^r)^T\cr -\beta_{[1]}^r&G_{(1)}\cr  \end{matrix} \right],\ee
where  $\alpha^r=\alpha\cdot \epsilon^r, \beta^r=\beta\cdot \epsilon^r$  and $\alpha^r_{[a]}, \beta^r_{[a]}$ denote $\alpha^r, \beta^r,$ respectively, omitting the $a$-th entry.
Then
\be\widetilde E^r=\Psi^o_m(k_1,\ldots,k_{m-1},-\bar\pi_m;\epsilon_1,\ldots,\epsilon_{m-1},\epsilon^{s};z_3,\ldots,z_{m-1})_{(2,m)}\ee
and, if we move the first row and column into the $(N-m+1)$-th positions, $ \widetilde G^r$ becomes
\be\Psi^o_{N-m+2}(\bar\pi_m,k_m,\ldots,k_N;\epsilon^{s},\epsilon_m,\ldots,\epsilon_N;x_{m+1},\ldots,x_{N-1})_{(1,N-m+2)}.\ee
In principle, in (\ref{asymPhi}), $\epsilon^r$ should run over the $d$ elements of a basis for the whole space, but we can restrict to a basis for the polarization states for $\bar\pi_m$, since the metric tensor is
$$\eta_{\mu\nu}=\sum_r\epsilon^r_\mu\epsilon^r_\nu+{1\over\bar\pi_m\cdot k}(k_\mu\bar\pi_{m\nu}+\bar\pi_{m\mu} k_\nu),$$
where $k$ is chosen so that $k\cdot\epsilon_r=k^2=0,k\cdot\bar\pi_m\ne0$, and the contributions from the last two terms vanish because 
\begin{align}
\Pf \Psi^o_m(&k_1,\ldots,k_{m-1},-\bar\pi_m;\epsilon_1,\ldots,\epsilon_{m-1},\bar\pi_m;z_3,\ldots,z_{m-1})_{(2,m)}\cr
&=
\Pf\Psi^o_{N-m+2}(\bar\pi_m,k_m,\ldots,k_N;\bar\pi_m,\epsilon_m,\ldots,\epsilon_N;x_{m+1},\ldots,x_{N-1})_{(1,N-m+2)}=0,\nonumber
\end{align}
using, in the first case, that $ \Psi^o_{m\,(2,m)}$ has null vector
$(1,z_3-1,\ldots,z_{m-1}-1,0,\dots,0,1)$ and, in the second case, that $\Psi^o_{N-m+2\,(1,N-m+2)}$
 has null vector $(1,x_{m+1},\ldots,x_{N-1},0,1,0,\dots,0)$.
 
 As $z_m\rightarrow 0$,
\be\prod_{a=2}^{N-1}(z_a-z_{a+1})\rightarrow z_{m-1}z_m^{N-m}\prod_{a=2}^{m-2}(z_a-z_{a+1})\prod_{a=m}^{N-1}(x_a-x_{a+1}),\ee
implying the desired factorization property as $z_m\rightarrow 0$,
\begin{align}
\Psi_N^o(k_1,\ldots,k_N&;\epsilon_1,\ldots,\epsilon_N;z_3,\ldots,z_{N-1})_{(2,N)}\cr
\sim\sum_s\Psi_m^o&(k_1,\ldots,k_{m-1},-\bar\pi_m;\epsilon_1,\ldots,\epsilon_{m-1},\epsilon^{s};z_3,\ldots,z_{m-1})_{(2,m)}
\cr
&\times\Psi_{N-m+2}^o(\bar\pi_m,k_m,\ldots,k_N;\epsilon^{s},\epsilon_m,\ldots,\epsilon_N;x_{m+1},\ldots,x_{N-1})_{(1,N-m-2)}.\cr
\end{align}
This, together with the corresponding result for the pole at $\zeta=\zeta_m^R$, which holds by symmetry, establishes the BCFW recurrence relation.

\section*{Acknowledgements}
We are grateful to Nima Arkani-Hamed and Song He for discussions.
LD thanks the Institute for Advanced Study at Princeton for its hospitality.
LD was partially supported by the U.S. Department of Energy, Grant No. 
DE-FG02-06ER-4141801, Task A, and PG was partially supported by NSF
grant No. PHY-1314311.
\vfill\eject
\appendix
\section{Properties of Pfaffians}
\label{Pfaff}
The Pfaffian $\Pf A$ is defined for an antisymmetric matrix $A=-A^T$ by $\det A=(\Pf A)^2$, subject to a choice of sign. If $\dim A$ is odd, $\Pf A=\det A=0$; if $\dim A$ is even, $\Pf A$ is a rational function of the coefficients of $A$. Explicitly, if $M=2N$, 
\be\Pf A = \sum_{\rho\in\frak S_{2N}} {1\over 2^NN! }\hbox{sign}\rho\,\prod_{i=1}^Na_{\rho(2i-1)\rho(2i)},\ee
where $\frak S_M$ denotes the group of permutations of $1,2,\ldots,M$.

We can write $\Pf A$ as a sum over the pairings $\varrho=\{(i_1,j_1), (i_2,j_2),\dots,(i_N,j_N)\}$  of the integers $1,2,\ldots, 2N-1,2N$. Denote the set of such pairings by $\frak P_N$. We can choose a unique description of the labeling of the pairing  by specifying that $i_a<j_a$ and $i_a<i_b$ if $1\leq a<b\leq N$. Given $\varrho\in\frak P_N$, define $\rho_\varrho\in \frak S_{2N}$ by $\rho_\varrho(2a-1)=i_a, \rho_\varrho(2a)=j_a, 1\leq a\leq N$, and set $\hbox{sign}\varrho=\hbox{sign}\rho_\varrho$. Then
\be\Pf A = \sum_{\varrho\in\frak P_N} \hbox{sign}\varrho\,\prod_{i=1}^Na_{i_aj_a.}\ee

 If $A_{(ij)}$ denotes the matrix obtained from $A$ by deleting the $i$-th and $j$-th rows and $i$-th and $j$-th columns. Then $A_{(ij)}$ is an antisymmetric matrix with $\dim A_{(ij)} = \dim A -2$. Then we have the expansion
\be\Pf A = \sum_{j=2}^M(-1)^j a_{1j}\Pf A_{(1j)} .\ee 
 
 In order to establish the factorization result in section \ref{GT}, we need  a result on the factorization of Pfaffians in suitable limits, which we prove here. First we establish a preliminary result, which may serve to elucidate the subsequent argument. 
   
 {\bf Preliminary Lemma.} Suppose $E,G$ are  antisymmetric matrices of dimensions of $m\times m$ and $n\times n$, respectively; $a,b$ are vectors of dimensions $m$ and $n$, respectively; and $F=ab^T,$ {\it i.e.} $F_{ij}=a_ib_j$. Let
\be\Phi=\left[\begin{matrix} E&F\cr -F^T&G\cr  \end{matrix}\right].\ee
Then, if $m+n$ is odd, $\Pf\Phi=0$; if $m+n,$ $m$ and $n$ are even, $\Pf\Phi =\Pf E\times \Pf G$; if $m+n$ is even, and $m$ and $n$ are odd, $\Pf\Phi =\Pf \widetilde E\times \Pf \widetilde G$, where
\be\widetilde E=\left[\begin{matrix} E&a^T\cr -a&0\cr  \end{matrix}\right],\qquad \widetilde G=\left[\begin{matrix}0&b^T\cr -b&G\cr  \end{matrix}\right].\ee

 {\it Proof:}
  Suppose $m+n$ even, $m+n=2N$.
\be\Pf \Phi = \sum_{\varrho\in\frak P_N} \hbox{sign}\varrho\,\prod_{a=1}^N \Phi_{i_aj_a},\ee
 where $\varrho=\{(i_1,j_1), (i_2,j_2),\dots,(i_N,j_N)\}$. Let $C_\varrho$ denote the number of $a$ for which $i_a\leq m$ and $j_a>m$. $m$ and $n$ are either both odd or both even and $C_\varrho$ has to be odd if $m,n$ are odd, and even if $m,n$ are even.
 
For each $\varrho\in\frak P_N$ with a given value of $C_\varrho>1$, we associate to $\varrho'\in\frak P_N$, which is obtained by replacing the pairs $(i_{c},j_{c}), (i_d,j_d)$, where $c,d$ are the largest values of $a$ such that $i_a\leq m$ and $j_a>m$, by the pairs $(i_{c},j_{d}), (i_d,j_c)$. Then $\hbox{sign}\varrho=-\hbox{sign}\varrho'$, but the value of the product
\be\prod_{a=1}^N \Phi_{i_aj_a.}\label{Prod}\ee
is the same for $\varrho'$ as for $\varrho$. Thus the contributions to the sum over $\varrho\in\frak P_N$ from those $\varrho$ with $C_\varrho>1$ cancel in pairs. 
Since $C_\varrho$ has the same parity as $m$ and $n$, all the nonzero contributions come from $C_\varrho=0$ if $m, n$ are even and from $C_\varrho=1$ if $m,n$ are odd.

If $m,n$ are even, each  pairing $\varrho\in\frak P_N$ making a nonzero contribution to $\Pf\Phi$ is the union of parings $\varrho_1\in\frak P_{\half m},$ of $1, 2, \ldots, m,$ and $\varrho_2\in\frak P_{\half n},$ of $m+1, 2, \ldots, m+n$.
So, the sum over such $\varrho\in\frak P_N$ is the product of sums over $\varrho_1\in\frak P_{\half m}$  and over $\varrho_2\in\frak P_{\half n},$ with $\hbox{sign}\varrho=\hbox{sign}\varrho_1\hbox{sign}\varrho_2.$ The product (\ref{Prod}) factors into two products corresponding to the pairings $\varrho_1$ and $\varrho_2$, respectively. Thus $\Pf\Phi =\Pf E\times \Pf G$. 

If $m,n$ are odd, each  $\varrho\in\frak P_N$ making a nonzero contribution to $\Pf\Phi$ involves just one pair $(i_0,j_0)$ with $i_0\leq m$ and $j_0>m$. The remaining pairs in $\varrho$ are the  union of parings $\varrho_1'\in\frak P_{\half (m-1)},$ of $\{ i: 1\leq i\leq m, i\ne i_0\},$ and $\varrho_2'\in\frak P_{\half (n-1)},$ of $\{ j: m< j\leq m+n, j\ne j_0\}$. Define the pairing $\varrho_1\in\frak P_{\half (m+1)},$ of $\{ i: 1\leq i\leq m;i_\ast\},$ by appending $(i_0,i_\ast)$ to $\varrho_1'$,
and the  pairing $\varrho_2\in\frak P_{\half (n+1)},$ of $\{ j:  j_\ast;m< j\leq m+n\}$, by appending $(j_\ast,j_0)$ to $\varrho_2'$. Then the product (\ref{Prod}) for $\varrho$ factors into products associated with $\varrho_1'$ and $\varrho_2'$, together with 
\be\Phi_{i_0j_0}=F_{i_0j_0}=a_{i_0}b_{j_0}=\widetilde E_{i_0i_\ast}\widetilde G_{j_\ast j_0},\ee
labeling the last row and column of $\widetilde E$ by $i_\ast$ and the first row and column of $\widetilde G$ by $j_\ast$, and, again, $\hbox{sign}\varrho=\hbox{sign}\varrho_1\hbox{sign}\varrho_2$. The sum over $\varrho\in\frak P_N$
corresponds to the product of sums over $\varrho_1\in\frak P_{\half (m+1)}$ and $\varrho_2\in\frak P_{\half (n+1)}$, yielding $\Pf\Phi =\Pf \widetilde E\times \Pf \widetilde G$.

Now we proceed to prove the result needed in section \ref{GT}.

 {\bf Factorization Lemma.} Again suppose $E,G$ are  antisymmetric matrices of dimensions of $m\times m$ and $n\times n$, respectively; but now suppose   that $a_i=(a_i^\alpha), 1\leq i\leq m,\; b_j=(b_j^\alpha), 1\leq j\leq n,$ are $d$-dimensional vectors and $F_{ij}=a_i\cdot b_j$ and consider the leading behavior of 
\be\Phi( z)=\left[\begin{matrix} E&F\cr -F^T& z^{-1}G\cr  \end{matrix}\right],\quad\hbox{as}\quad  z\rightarrow 0.\ee
If $m+n$ is odd, $\Pf\Phi( z)=0$; if $m+n, m, n$ are even, $\Pf\Phi( z) \sim z^{-\half n}\Pf E\times \Pf G$, as $ z\rightarrow 0$; if $m+n$ is even, and $m$ and $n$ are odd, 
\be\Pf\Phi( z) \sim z^{-\half (n-1)}\sum_\alpha\Pf \widetilde E^\alpha\times \Pf \widetilde G^\alpha\quad\hbox{as }  z\rightarrow 0,\ee
 where
\be \widetilde E^\alpha=\left[\begin{matrix} E&(a^\alpha)^T\cr -a^\alpha&0\cr  \end{matrix}\right],\qquad \widetilde G^\alpha=\left[\begin{matrix} 0&(b^\alpha)^T\cr -b^\alpha&G\cr  \end{matrix}\right].\ee

 {\it Proof:} Suppose $m+n$ even, $m+n=2N$. With $\varrho=\{(i_1,j_1), (i_2,j_2),\dots,(i_N,j_N)\}$, the contribution 
 \be \hbox{sign}\varrho\,\prod_{a=1}^N \Phi(z)_{i_aj_a},\label{prho}\ee
to $\Pf\Phi( z)$ is ${\cal O}( z^{-M_\varrho})$, where $M_\varrho$ is the number of $a$ for which $i_a>m$, for then $\Phi( z)_{i_aj_a}={\cal O}( z^{-1}),$ but $\Phi( z)_{i_aj_a}={\cal O}(1)$ otherwise. If $C_\varrho$ again denotes the number of $a$ for which $i_a\leq m$ and $j_a>m$, $M_\varrho=\half(n-C_\varrho)$, so the behavior of $\Pf\Phi( z)$ as $  z\rightarrow 0$ is dominated by terms with $C_\varrho$ as small as possible, {\it i.e.} $C_\varrho=0$ for $m,n$ even and $C_\varrho=1$ for $m,n$ odd. So, reasoning as in Lemma 1, if $m,n$ are even, the leading contributions come from $\varrho\in\frak P_N$ which are the union of parings $\varrho_1\in\frak P_{\half m},$ of $1, 2, \ldots, m,$ and $\varrho_2\in\frak P_{\half n},$ of $m+1, 2, \ldots, m+n,$ and $\Pf\Phi( z) \sim z^{-\half n}\Pf E\times \Pf G$, as $ z\rightarrow 0$.

If $m,n$ are odd, each  $\varrho\in\frak P_N$ making a leading contribution to $\Pf\Phi$ involves just one pair $(i_0,j_0)$ with $i_0\leq m$ and $j_0>m$. As in Lemma 1, the remaining pairs in $\varrho$ are the  union of parings $\varrho_1'\in\frak P_{\half (m-1)},$ of $\{ i: 1\leq i\leq m, i\ne i_0\},$ and $\varrho_2'\in\frak P_{\half (n-1)},$ of $\{ j: m< j\leq m+n, j\ne j_0\}$. So again, define the pairing $\varrho_1\in\frak P_{\half (m+1)},$ of $\{ i: 1\leq i\leq m;i_\ast\},$ by appending $(i_0,i_\ast)$ to $\varrho_1'$,
and the  pairing $\varrho_2\in\frak P_{\half (n+1)},$ of $\{ j:  j_\ast;m< j\leq m+n\}$, by appending $(j_\ast,j_0)$ to $\varrho_2'$. Then the product (\ref{prho}) for $\varrho$ factors into products associated with $\varrho_1'$ and $\varrho_2'$, together with 
\be\Phi_{i_0j_0}=F_{i_0j_0}=\sum_\alpha a_{i_0}^\alpha b_{j_0}^\alpha =\sum_\alpha \widetilde E_{i_0i_\ast}^\alpha \widetilde G_{j_\ast j_0}^\alpha ,\ee
labeling the last row and column of $\widetilde E$ by $i_\ast$ and the first row and column of $\widetilde G$ by $j_\ast$, and, again, $\hbox{sign}\varrho=\hbox{sign}\varrho_1\hbox{sign}\varrho_2$. The sum over $\varrho\in\frak P_N$
corresponds to the product of sums over $\varrho_1\in\frak P_{\half (m+1)}$ and $\varrho_2\in\frak P_{\half (n+1)}$, yielding 
\be\Pf\Phi( z) \sim z^{-\half (n-1)}\sum_\alpha\Pf \widetilde E^\alpha\times \Pf \widetilde G^\alpha\quad\hbox{as }  z\rightarrow 0.\ee 

\vfil\eject
\section{Massive ${\boldsymbol\phi}^{\bold 3}$  theory.}
\label{massive}
In this appendix, we show how the results for tree amplitudes in massless $\phi^3$ theory, established in sections \ref{SEqs} to \ref{phi33}, can be extended to the massive case by replacing $\hat f_a$, defined by 
(\ref{SE}), by $\breve f_a$, defined as in (\ref{MSE}),
\be
\check f_a(z,k)=\sum_{b\in A\atop b\ne a}{k_a\cdot k_b\over z_a-z_b}-{a_0\over 2(z_a-z_{a+1})}-{a_0\over 2(z_a-z_{a-1})},\qquad a\in A.\label{MSE2}
\ee

As in section \ref{SEqs}, suppose $z_a\rightarrow z_S$ for $z\in S\subset A$, with  $z_a= z_S +\epsilon x_a+\O(\epsilon^2)$, as $\epsilon\rightarrow 0$, and $z_a\not\rightarrow z_S$ for $z\notin S$. Then, in place of (\ref{SES}) and (\ref{SES1}), we have 
\be
\breve f_a(z,k)={1\over\epsilon}\breve g_a(x,k)\left[1+\O(\epsilon)\right],\quad\breve g_a(x,k)=\sum_{b\in S\atop b\ne a}{k_a\cdot k_b\over  x_a- x_b}
-\sum_{b\in S\atop b= a\pm1}{a_0\over  x_a- x_b},\quad a\in S, \label{MSES}
\ee
\be
\breve f_a(z,k)=\sum_{b\in S\atop }{k_a\cdot k_b\over z_a-z_S}-\sum_{b\in S\atop b=a\pm 1}{a_0\over z_a-z_S}+\sum_{b\notin S\atop b\ne a}{k_a\cdot k_b\over z_a-z_b}-\sum_{b\notin S\atop b= a\pm1}{a_0\over z_a-z_b}+\O(\epsilon),\quad a\notin S.\label{MSES1}
\ee
Then, as in (\ref{sumg}),
\be
\sum_{a\in S} \breve g_a(x,k)=\sum_{a,b\in S\atop b\ne a}{k_a\cdot k_b\over  x_a- x_b}
-\sum_{a,b\in S\atop b= a\pm1}{a_0\over  x_a- x_b}=0,\label{Msumg}
\ee
implying that if $\breve g_a=0$ for $a\in S, a\ne s$, where $s$ is some particular element of $S$, then it follows that $\breve g_s(x)=0$ as well.

Also, as in (\ref{SES2}),
\begin{align}
\sum_{a\in S}(x_a-x_r)\breve g_a(x,k)&=\sum_{a,b\in S\atop b\ne a}{x_a-x_r\over  x_a- x_b}k_a\cdot k_b-\sum_{a,b\in S\atop b= a\pm1}{x_a-x_r\over  x_a- x_b}a_0\
=\half k_S^2+na_0,\qquad  k_S=\sum_{b\in S\atop }k_b,\label{MSES2}
\end{align}
where $2n$ is the number of $a$ for which $a+1\notin S$ plus the number of $a$ for which $a-1\notin S$. Then, if $\breve g_a(x,k)=0, a\in S, a\ne r,s$, as in (\ref{grs}), and noting that $m^2=-2a_0$, we have that
\be
(x_s-x_r)\breve g_s(x,k)=\half (k_S^2 - nm^2)=(x_r-x_s)\breve g_r(x,k),\label{Mgrs}
\ee
and, if additionally $g_r(x,k)=0$, then $k_S^2=nm^2$. If $S$ is consecutive, and not the whole of $A$, then $n=1$, and this becomes the mass-shell condition.

Again, imposing the equations $\breve f_a(z,k)=0,$ for all $a\in A,$ and taking the limit $\epsilon\rightarrow0$, we obtain, as in (\ref{kS2}), a factorization into two sets of equations,
\be
\sum_{b\in S\atop b\ne a}{k_a\cdot k_b\over  x_a- x_b}
-\sum_{b\in S\atop b= a\pm1}{a_0\over  x_a- x_b}=0,\quad a\in S, \label{MkS21}
\ee
\be
{k_S\cdot k_b\over z_a-z_S}-\sum_{b\in S\atop b=a\pm 1}{a_0\over z_a-z_S}+\sum_{b\notin S\atop b\ne a}{k_a\cdot k_b\over z_a-z_b}-\sum_{b\notin S\atop b= a\pm1}{a_0\over z_a-z_b}=0,\quad a\notin S,
 \label{MkS22}
\ee
 and, provided that $S$ is consecutive, the first set of equations are the modified CHY equations for the momenta $(k_a, a\in S; -k_S)$, with associated variables $(x_a, a\in S; \infty)$, and the second set are the equations for the momenta $(k_S;k_a, a\notin S)$, with associated variables $(z_S;z_a, a\notin S)$.
 
 The results (\ref{MSES}) to (\ref{MkS22}) make it straightforward to adapt the analysis of sections \ref{phi31}, \ref{phi32} and \ref{phi33} to the massive case. If we replace $\hat f_a(z,k)$ in (\ref{ampphi1}) by $\breve f_a(z,k)$to obtain 
\be
\A_N^{\phi,m}={1\over (-2)^{N-3}}\oint_\O {\prod_{a\in A\atop }}'{1\over\breve f_a(z,k)}\prod_{a\in A}{dz_a\over (z_a-z_{a+1})^2}\bigg/d\omega\label{ampphi1M}
\ee
the effect in general is to replace $k_a\cdot k_b$ by $k_a\cdot k_b-a_0$ whenever $b=a\pm1$, so that, for example, (\ref{amp4}) becomes
\begin{align}
\A_4^{\phi,m}&={1\over 2}\oint_\O{dz\over (1-z)z[(k_3\cdot k_2-a_0)z+(k_3\cdot k_4-a_0)(z-1)]} ,\cr
&={k_3\cdot (k_2+ k_4)-2a_0\over 2(k_3\cdot k_2-a_0)(k_3\cdot k_4-a_0)}={1\over s+m^2}+{1\over t+m^2}.
\label{Mamp4}
\end{align}
Further the effect is to replace $s_{ab}=(k_a+k_b)^2$ by $s_{ab}-m^2=(k_a+k_b)^2-m^2$ and 
$s_{abc}=(k_a+k_b+k_c)^2$ by $s_{abc}-m^2=(k_a+k_b+k_c)^2-m^2$ in (\ref{R1}) to (\ref{R4}) [Note that $a,b$ and $a,b,c$ are consecutive where they occur in $s_{ab}$ and $s_{abc}$.], giving the correct result for $\A_5^{\phi,m}$. Because of (\ref{MSES2}) and (\ref{Mgrs}), the effect is to replace $s_m$ with $s_m - m^2$ and $\bar s_m$ with $\bar s_m - m^2$  throughout sections \ref{phi32} and \ref{phi33}, so that those sections provide inductive proofs of that $\A_N^{\phi,m}$ provides the correct $N$-point tree amplitudes for massive $\phi^3$ theory.

\vfill\eject

\vfill\eject

\singlespacing


\providecommand{\bysame}{\leavevmode\hbox to3em{\hrulefill}\thinspace}
\providecommand{\MR}{\relax\ifhmode\unskip\space\fi MR }
\providecommand{\MRhref}[2]
{
}
\providecommand{\href}[2]{#2}

\end{document}